\documentclass[sigconf]{acmart}
\usepackage{hyperref}

\setcopyright{none}
\renewcommand\footnotetextcopyrightpermission[1]{} % removes footnote with conference information in first column
\AtBeginDocument{%
  \providecommand\BibTeX{{%
    \normalfont B\kern-0.5em{\scshape i\kern-0.25em b}\kern-0.8em\TeX}}}

\acmConference[arXiv]{XXX}{May 2023}{Berkeley, CA, USA}

\begin{document}

\title{NFT.mine: An xDeepFM-based Recommender System for Non-fungible Token (NFT) Buyers}

\author{Shuwei Li}
\affiliation{%
  \institution{University of California-Berkeley}
  \city{Berkeley}
  \state{California}
  \country{USA}}
 \email{shuwei@berkeley.edu}
 
 \author{Yucheng Jin}
\affiliation{%
  \institution{University of California-Berkeley}
  \city{Berkeley}
  \state{California}
  \country{USA}}
 \email{yuchengjin@berkeley.edu}
 
 \author{Pin-Lun Hsu}
\affiliation{%
  \institution{University of California-Berkeley}
  \city{Berkeley}
  \state{California}
  \country{USA}}
 \email{byron\_hsu@berkeley.edu}
 
 \author{Ya-Sin Luo}
\affiliation{%
  \institution{University of California-Berkeley}
  \city{Berkeley}
  \state{California}
  \country{USA}}
 \email{yasin\_luo@berkeley.edu}

\renewcommand{\shortauthors}{Shuwei Li et al.}

\begin{abstract}
Non-fungible token (NFT) is a tradable unit of data stored on the blockchain which can be associated with some digital asset as a certification of ownership. The past several years have witnessed the exponential growth of the NFT market. In 2021, the NFT market reached its peak with more than \$40 billion trades. Despite the booming NFT market, most NFT-related studies focus on its technical aspect, such as standards, protocols, and security, while our study aims at developing a pioneering recommender system for NFT buyers. In this paper, we introduce an extreme deep factorization machine (xDeepFM)-based recommender system, NFT.mine\footnote{Source code available at  \color{blue}{\url{https://github.com/wallerli/NFT.mine}}}, which achieves real-time data collection, data cleaning, feature extraction, training, and inference. We used data from OpenSea, the most influential NFT trading platform, to testify the performance of NFT.mine. As a result, experiments showed that compared to traditional models such as logistic regression, naive Bayes, random forest, etc., NFT.mine outperforms them with higher AUC and lower cross entropy loss and outputs personalized recommendations for NFT buyers.
\end{abstract}




\begin{teaserfigure}
  \centering
  \includegraphics[width=15.5cm]{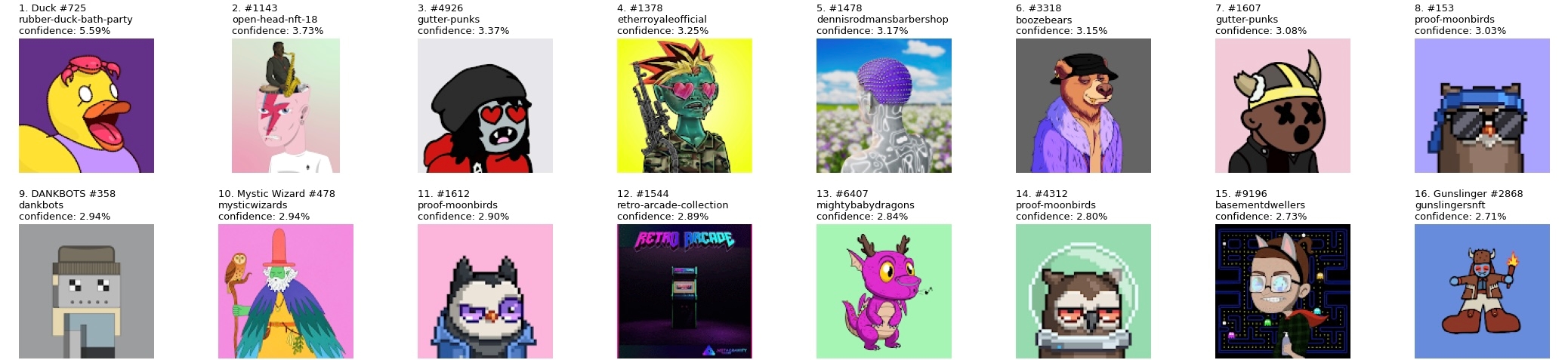}
  \caption{NFT.mine outputs personalized recommendations for an NFT buyer}
\end{teaserfigure}

\settopmatter{printfolios=true}
\maketitle

\section{Introduction}

\subsection{Overview of Non-fungible Token (NFT)}
Non-fungible token (NFT) is a unit of data stored on the blockchain \cite{wilson2021prospecting}, which can be used as a certification of ownership associated with some digital asset because of its cryptographic characteristics, including uniqueness, indivisibility, transferability, and encipherment protection \cite{wang2021non}. 

Blockchain is the core technology of NFT, which can be regarded as one kind of distributed database system where transaction information is stored with encoding and each transaction can be validated by all blockchain users \cite{nofer2017blockchain}.

Compared to other blockchain-based cryptocurrencies such as Bitcoin (BTC), Dogecoin (DOGE), Ethereum (ETH), etc. \cite{blockchaincurrencies}, NFT is the most ideal choice to identify the uniqueness of digital commodities, since classical cryptocurrencies are not distinguishable \cite{nakamoto2008bitcoin}. For example, two BTCs can be completely identical and exchangeable. In contrast, the intrinsic non-fungible attribute of NFT enables its owner to prove the ownership in a transparent form, as every historical transaction is retrievable \cite{wood2014ethereum}.

Besides blockchain, smart contract and encoding are also crucial technologies for NFT. The idea of smart contract was first introduced for efficient digital negotiations and was widely implemented in Ethereum systems \cite{szabo1996smart}. It provides a standard decentralized method for digital currency exchange. Based on consistently shared parameters and instructions across the distributed nodes, smart contract makes transactions transparent across multi-parties, which forms the foundation of NFT trading \cite{evans2019cryptokitties}.

Encoding is widely used to realize blockchains and smart contracts. By converting transaction information into compressed and encrypted data, transaction history can only be decoded and validated by owners with private keys (e.g. hex-value keys) \cite{wood2014ethereum}.

NFT has a lot of advantages because of its decentralized architecture, \cite{wang2021non} and these advantages make NFT an emerging and promising technology for many real-world applications. For example, traditional online ticketing usually needs the help of a trusted third party, such as TicketMaster \cite{zynda2004ticketmaster}, while NFT-based ticketing is an efficient and safe solution for online ticketing without any trusted third party \cite{wang2021non}. Metaverse is a new concept developed recently, where people are connected in a virtual world that operates like the real society \cite{mystakidis2022metaverse}. In metaverse, NFT is a tool for transactions, some well-known NFT-empowered metaverses are Sandbox \footnote{\color{blue}{\url{https://sandboxvr.com/}}} and Decentraland \footnote{\color{blue}{\url{https://decentraland.org/}}}. In addition to online ticketing and metaverse, NFT is also widely used for digital assets trading on online platforms, such as OpenSea \footnote{\color{blue}{\url{https://opensea.io/}}}, which we will mention later.

\subsection{The NFT Market}
In 2021, the NFT market reached a historical record-break high, as nearly \$41 billion worth of cryptocurrency trades were made \cite{Bloomberg}. Looking back on the development of NFT, the first noticeable and popular NFT-based trading example is CryptoKitties, a game that allows its players to buy, breed, and sell virtual pets on Ethereum \cite{Quartz}. However, for nearly 2 years since CryptoKitties, the NFT market didn't see much expansion. It was not until July 2020, an artist called Beeple auctioned an NFT of his creation at Christie's for more than \$69 million \cite{Christie}, the NFT market experienced an explosive growth. 

The items traded on the NFT market are called collections, which can be clustered into 6 categories: art, collectible, games, metaverse, other, and utility \cite{nadini2021mapping}. Based on these 6 categories, Nadini et al. \cite{nadini2021mapping} analyzed each category's share of volume and transactions with respect to time, distribution of NFT prices across different categories, and frequencies of individual assets exchanges. As a result, the NFT market is continuously growing, pricing of NFT is complicated as NFT prices fluctuate a lot, and NFT buyers form clear clusters. Therefore, a recommender system for NFT buyers is essential and profitable, because it can help buyers to locate the most attractive NFTs for them at reasonable prices.

\subsection{Recommender System}

A recommender system is a subclass of information filtering system \cite{ricci2011introduction} that receives user information such as ratings, preferences, search results, and generates personalized recommendations to a collection of users for commodities that might attract their eyes \cite{melville2010recommender}. Collaborative filtering, content-based recommendation, hybrid recommendation are three most commonly used models for a recommender system \cite{adomavicius2005toward}.

In recent years, deep learning played an important role in recommender systems, and it was proved to be a feasible choice to improve recommendation performance \cite{zhang2019deep}. In this study, we introduce NFT.mine that uses deep learning for NFT recommendation. NFT.mine receives trading records collected in real-time from OpenSea platform, interprets NFT features by xDeepFM-based architect, and outputs personalized recommendations for each NFT buyer in the market. 

\section{Related Work}
In this section, we introduce some previous studies on NFT and recommender systems that provide us with incisive insights into this study. 

Many studies focus on the methodologies for NFT market analysis. One example is mentioned above in Section 1.2, which is Nadini et al.'s  \cite{nadini2021mapping} study based on NFT categorization and market analysis as a function of time. Similar to Nadini et al.'s study, Wang et al. \cite{wang2021non} used primary-sales and secondary-sales data to investigate the NFT market's activity. Other studies combine the analysis of the NFT market with cryptocurrencies. For example, Michael Dowling \cite{dowling2021non} defined a spillover index and found the influence of volatility transmission effect between cryptocurrencies and NFT trading is not significant, while his wavelet coherence analysis suggests a correlation between cryptocurrencies and NFTs. Lennart Ante \cite{ante2021non} proposed a vector autoregressive (VAR) framework to understand the interrelationships between NFT sales, NFT users, and pricings of cryptocurrencies; as a result, Ante concluded that the NFT market is dependent on the cryptocurrency market.

NFT pricing and sales prediction is another heated topic. For example, Michael Dowling \cite{dowling2022fertile} collected 4,936 secondary market trades data in Decentraland (a blockchain virtual world) and used automatic variance ratio (AVR) test, automatic portmanteau (AP) test, and Domínguez and Lobato (DL) \cite{dominguez2003testing} consistent test to price NFTs. Nadini et al. \cite{nadini2021mapping} constructed a linear regression model to estimate the primary-sales and secondary-sales of NFTs. Besides the simplest linear regression model, other more complicated models can also be used to price NFTs. An example is Schnoering and Inzirillo \cite{schnoering2022constructing}, they built a multiplicative pricing model based on the scale price, assets traits, scarcity of assets, and global state of the NFT market. As a result, the multiplicative pricing model can be applied for diagnostic tests, dynamics analysis, and performance evaluation of the NFT market.

Based on studies about NFT market analysis and NFT pricing and sales prediction, we summarized methodologies for data preprocessing and exploratory data analysis (EDA) and implemented these methodologies in our study. The data preprocessing and EDA parts will be introduced in detail in Section 3. 

Studies that focus particularly on recommender systems for NFT buyers are rare, but studies that investigate recommender systems with deep learning techniques can provide us with some incisive information. Zhang et al. \cite{zhang2019deep} summarized two deep learning-based recommendation categories, recommendation with neural building blocks and recommendation with deep hybrid models. For the first category, different neural networks are used for different purposes. For example, MLP learns feature extraction well, which was utilized by Covington et al. \cite{covington2016deep} for YouTube recommendation and Alashkar et al. \cite{alashkar2017examples} for makeup recommendation. For the second category, different deep learning techniques are combined together to build more powerful recommender systems. For example, our proposed recommender system, NFT.mine, is based on extreme deep factorization machine (xDeepFM) proposed by Lian et al. \cite{lian2018xdeepfm}.

\section{Dataset Preparation}

\subsection{Dataset Collection}
We conducted comprehensive research on the NFT market and concluded that the most popular NFT trading platform that is open to data collection is OpenSea. OpenSea API is open for developers and allows developers to get asset, event, account, and collection information.

Based on OpenSea API, we developed a real-time Python scraper that retrieves each NFT transaction event at each timestamp, then saved the collected data as a JSON file for data cleaning and EDA.

\begin{table*}
  \caption{Representative features of the OpenSea dataset}
  \label{tab:commands}
  \begin{tabular}{clll}
    \toprule
    Feature & Meaning & \# of Unique Values & Most Frequent Value\\
    \midrule
    \texttt{asset\_id} & Asset ID & 309,383 & 381803879 (550 times)\\
    \texttt{num\_sales}& Number of sales associated with the asset & 23 & 1 (175,180 times)\\
    \texttt{asset\_image\_url}& Asset image URL & 232,517 & https://lh3.googleusercontent.com/... (3,984 times)\\
    \texttt{asset\_name}& Asset name & 255,065 & Life (4,233 times)\\    
     \texttt{asset\_address}& Asset contract address & 4,595 & 0x582048c4077a34e7c3799962f1f8c... (8,633 times)\\    
    \texttt{collection\_slug}& Collection category & 4,669 & galverse (8,633 times)\\    
    \texttt{created\_date}& Timestamp of the transaction & 368,693 & 2022-04-16T02:32:03.169797 (2 times)\\  
    \texttt{event\_type}& Status of the transaction & 2 & successful (363,070 times)\\ 
    \bottomrule
  \end{tabular}
\end{table*}

The dataset used in this paper contains 396,707 NFT transaction records from OpenSea database with 185 features from 12 April 2022 to 17 April 2022. 

Table 1 lists some important features of the OpenSea dataset, which includes feature meaning, number of unique values, and the most frequent value of each feature.

\subsection{Data Cleaning}

After data collection, we did data cleaning. \textit{successful} and \noindent\textit{bid\_withdrawn} are two event types (\texttt{event\_type}) from which we could get sufficient features. \textit{successful} means a transaction was completed successfully, \textit{bid\_withdrawn} means the NFT buyer withdrew his/her bid and the transaction was canceled. We set the label of both event types to 1 as the buyer shows transaction interest. We then generated all combinations of user and NFT asset pairs and sampled data associated with label 0, which means a NFT buyer shows no interest in buying a particular NFT.

Besides target variable selection, we did date-time alignment, empty rate calculation, and obtained each feature's most popular values. We conducted date-time alignment to make sure all transaction timestamps fall between 2022-04-12 15:00 to 2022-04-17 21:00. Afterwards, we removed columns that have a large proportion of empty entries by calculating the empty rate ($\textit{empty rate}=\frac{\textit{number of empty entries}}{\textit{total entries}}$) of each column. We set a threshold of 0.25 and removed features with an empty rate larger than 0.25 from the set of training features. After this step, we obtained the number of unique values of each feature and the most frequent values to have a general idea of the training features.

\subsection{Exploratory Data Analysis (EDA)}

\subsubsection{Univariate Analysis} Exploratory data analysis (EDA) in this study contains four parts. For the first part, univariate analysis, we focused on features such as transaction status, payment token, user loyalty, number of sales associated with an asset, etc., to find out patterns in the OpenSea dataset. 

\begin{figure}[h]
  \centering
  \includegraphics[width=8cm]{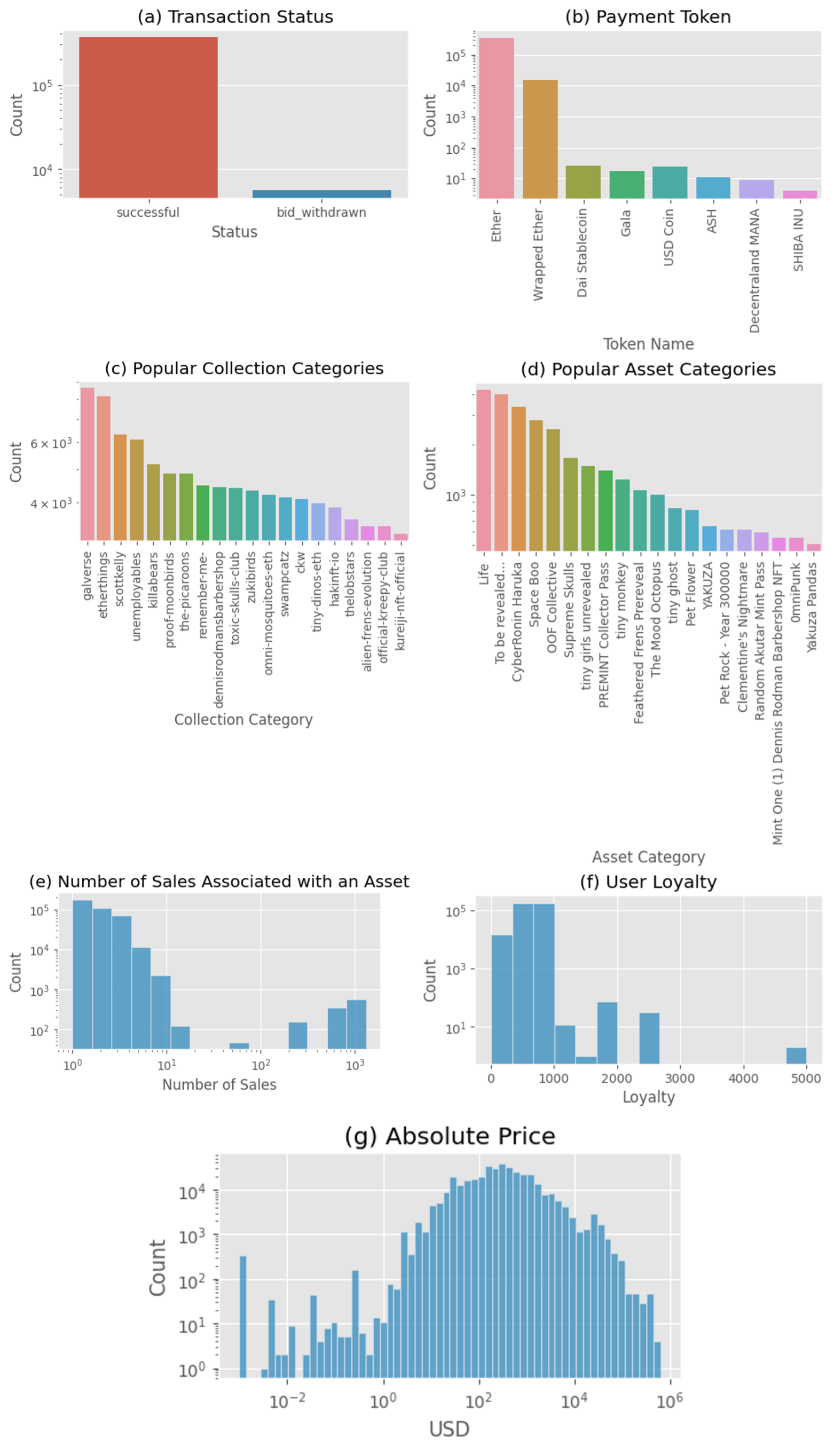}
  \caption{Univariate analysis}
  \label{fig:uni}
\end{figure}

As a result, we found that most transactions are labeled \textit{successful} (Fig.\ref{fig:uni}.a) and the top two payment tokens are \textit{Ether} and \textit{Wrapped Ether} (Fig.\ref{fig:uni}.b); the most popular collection categories are \textit{galverse}, \textit{etherthings}, \textit{scottkelly}, etc. (Fig.\ref{fig:uni}.c) and the most popular asset categories are \textit{Life}, \textit{To be revealed}, \textit{CyberRonin Haruka}, etc. (Fig.\ref{fig:uni}.d); for each transaction, the number of sales associated with the asset is around 1 to 10 (Fig.\ref{fig:uni}.e) and the user loyalty falls between 0 to 1,000 (Fig.\ref{fig:uni}.f); in addition, we visualized the distribution of absolute NFT prices in USD (Fig.\ref{fig:uni}.g) and found that the majority of sales are at a price from 10 USD to 10,000 USD.

\subsubsection{Correlation Analysis}  We conducted pairwise correlation analysis for feature selection.

\begin{figure}[h]
  \centering
  \includegraphics[width=8cm]{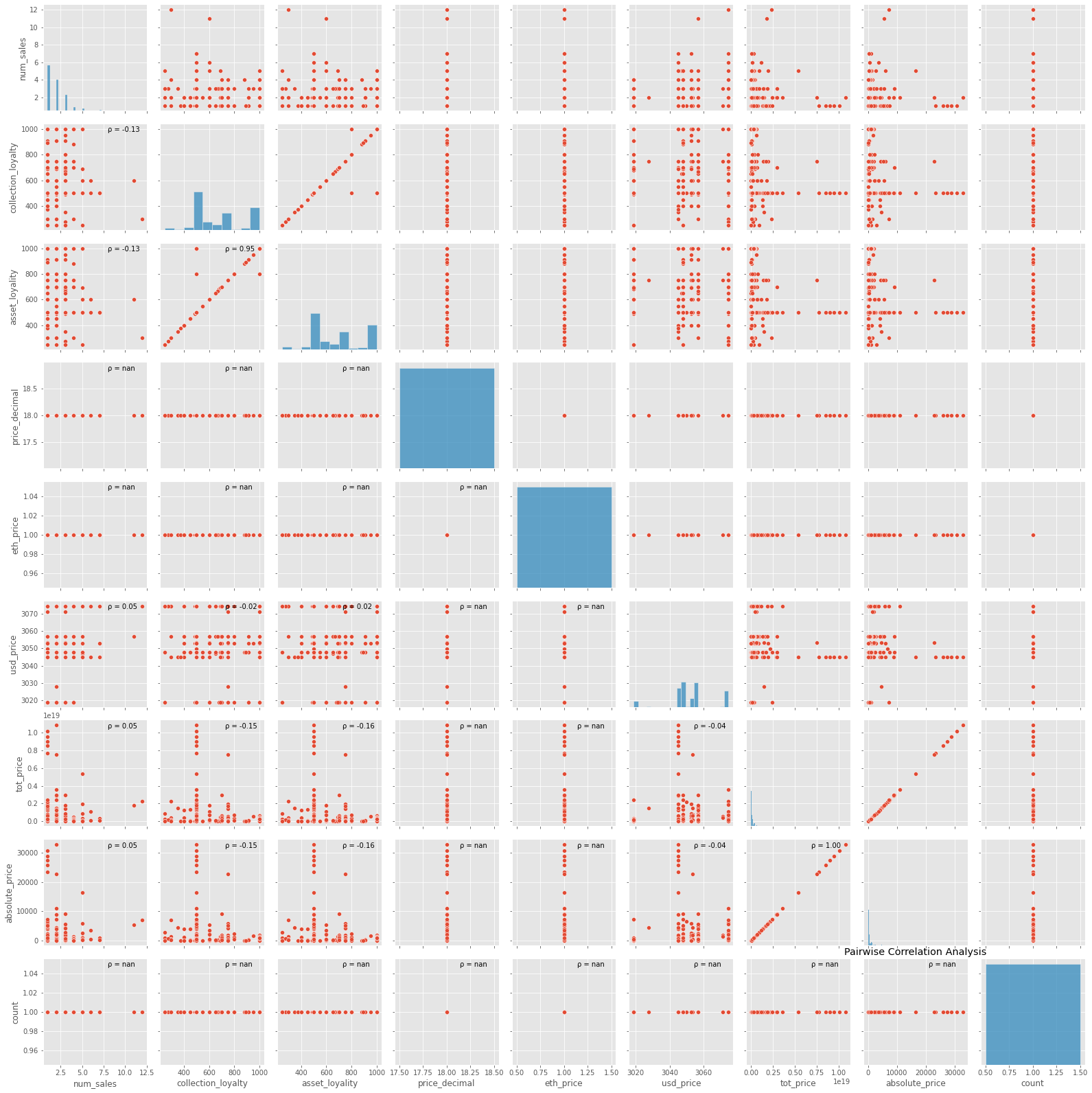}
  \caption{Pairwise correlation analysis}
  \label{fig:pair}
\end{figure}

As a result, we picked 22 features and found that \texttt{asset\_loyalty} highly correlates with \texttt{collection\_loyalty} and \texttt{abosolute\_price} and \texttt{total\_price} also have a strong positive correlation (Fig.\ref{fig:pair}). These highly correlated features bring no additional information, so we dropped some redundant features to reduce the model complexity.

\subsubsection{Bivariate Analysis} 
\begin{figure}[h]
  \centering
  \includegraphics[width=8.5cm]{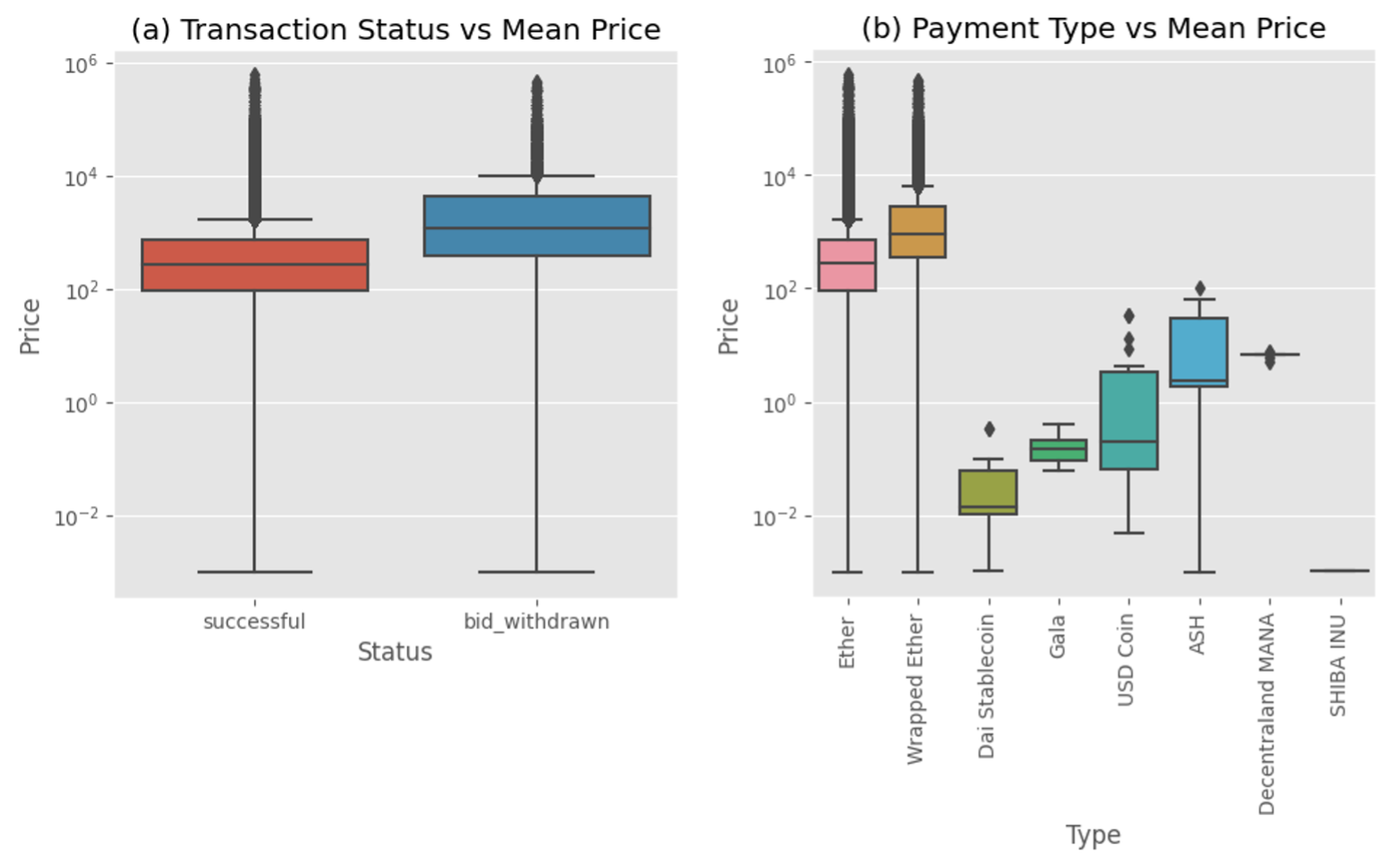}
  \caption{Bivariate analysis}
  \label{fig:bi}
\end{figure}

We visualized the relationships between price and transaction status (Fig\ref{fig:bi}.a) and price and payment type (Fig\ref{fig:bi}.b). From the box plots, we can tell that \textit{bid\_withdrawn} transactions correspond to a higher mean price than \textit{successful} ones, while different payment types correspond to different mean prices.

\subsubsection{Multivariate analysis} 
\begin{figure}[h]
  \centering
  \includegraphics[width=8cm]{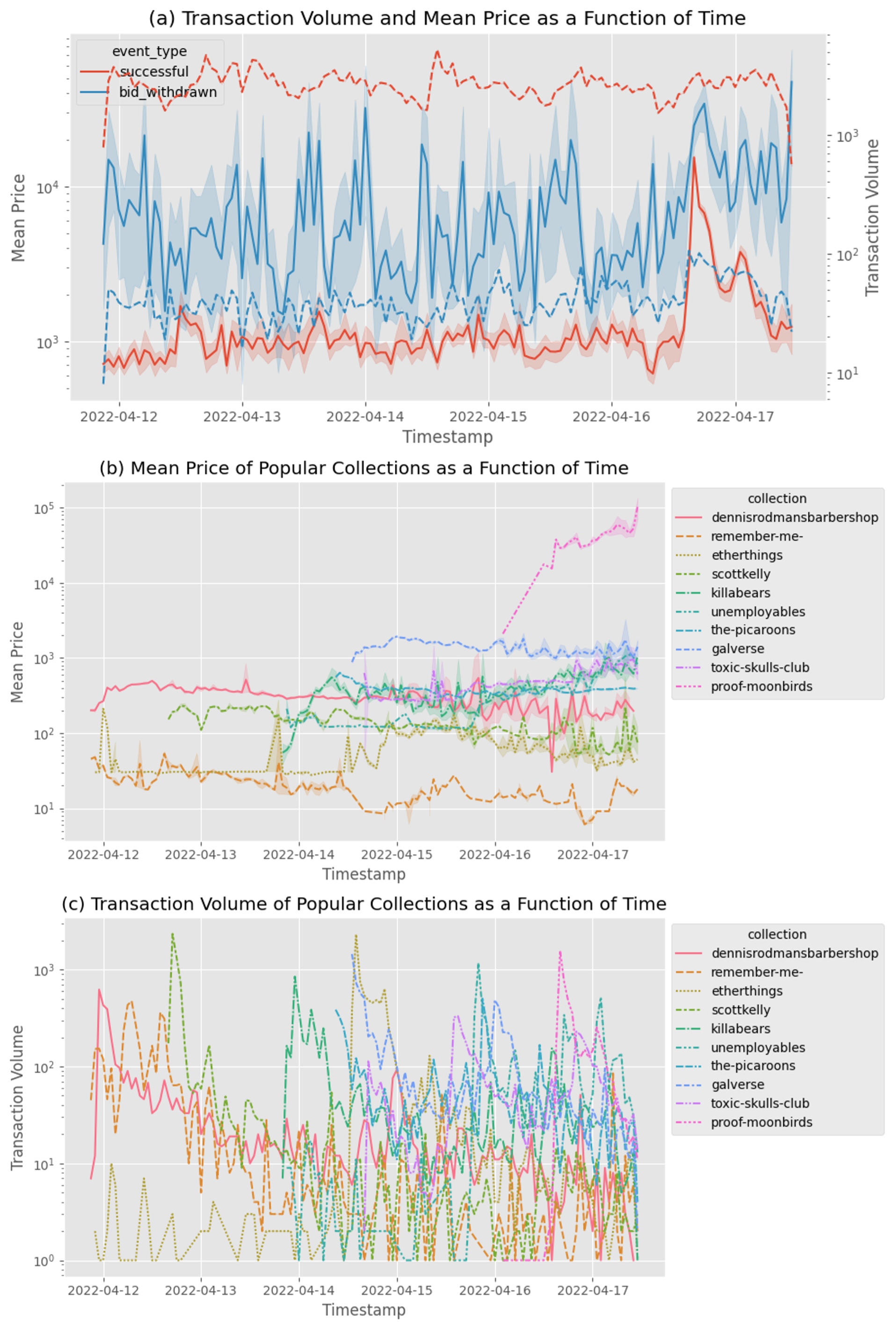}
  \caption{Market trend as a function of time}
  \label{fig:market}
\end{figure}

We also conducted multivariate analysis and plotted the market trend as a function of time (Fig.\ref{fig:market}).

\section{Approach}

This section introduces our proposed xDeepFM-based recommender system, NFT.mine. We first compare the performance of NFT.mine with other baseline models by various metrics. Then we explain the architecture and functionalities of NFT.mine. Because xDeepFM is capable of interpreting deep factorization of the data features, we can use it to develop the first-ever high-performing recommender system for NFT buyers, which is a pioneering work in this area.

% How and why does your solution improve on the related works you discussed?

\subsection{Baseline Models}

\subsubsection{Logistic Regression (LR)} LR predicts the probability of an event by sigmoid function with a linear combination of inputs \cite{wright1995logistic}. We implemented LR using \texttt{LogisticRegression} from \texttt{sklearn} library.

\subsubsection{Naive Bayes (NB)} NB assumes independence between input features and can be used for probability estimation \cite{lowd2005naive}. We implemented NB using \texttt{GaussianNB} from \texttt{sklearn} library.

\subsubsection{Random Forest (RF)} RF is an ensemble learning method for classification \cite{belgiu2016random}. We implemented RF using \texttt{RandomForestClassifier} from \texttt{sklearn} library.

\subsection{Evaluation Metrics}

\subsubsection{Area under the ROC Curve (AUC)} AUC measures the area underneath the ROC curve, ranging from 0 to 1, higher AUC value represents better performance.

\subsubsection{Cross Entropy Loss (Logloss)} Logloss measures the difference between predicted labels and true labels, lower Logloss value represents better performance.

\subsection{NFT.mine}

\begin{figure*}[h]
  \centering
  \includegraphics[width=10cm]{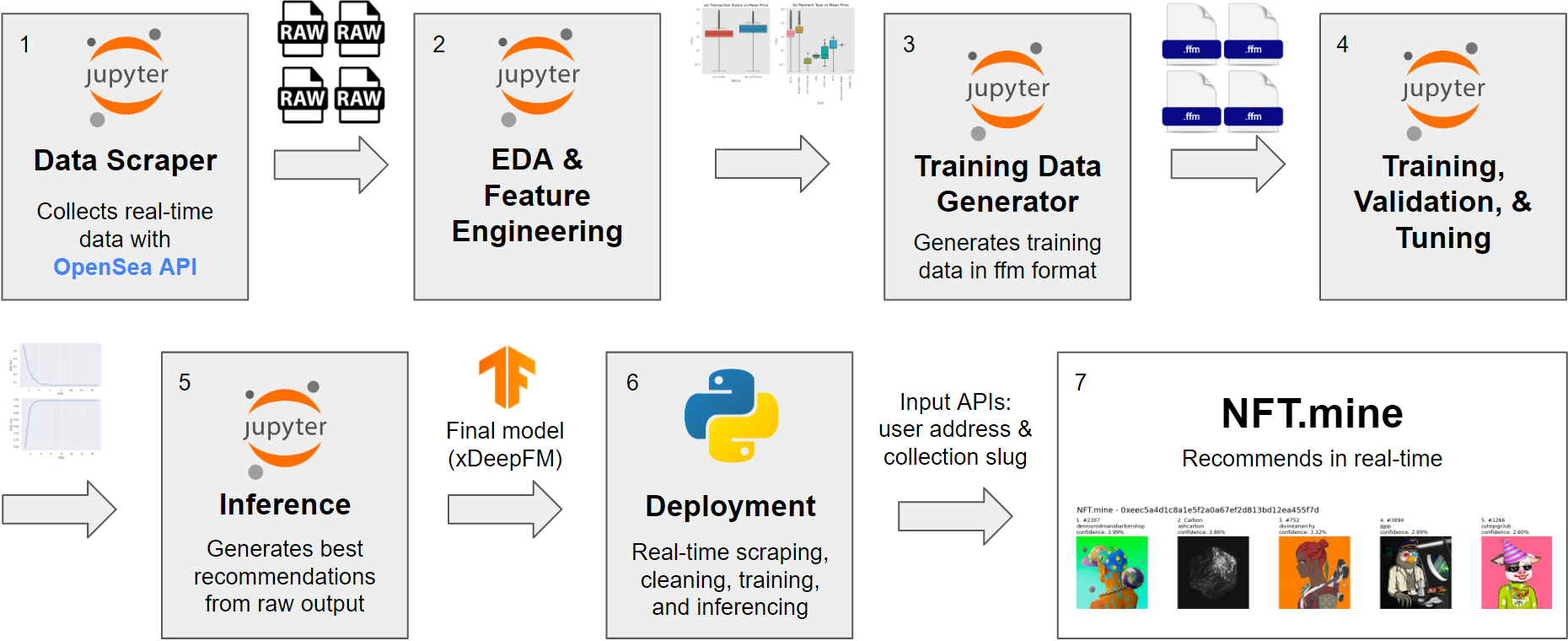}
  \caption{NFT.mine architecture}
  \label{fig:arch}
\end{figure*}

NFT.mine is an end-to-end recommender system, including real-time data collection, data analysis, feature selection, model training, and model inference (Fig.\ref{fig:arch}). There are five modules in NFT.mine, including Python scrapper, EDA module, dataset module, server module, and xDeepFM model.

We built NFT.mine based on \cite{recommender}. We integrated significant modifications, including metric collection, hyperparameter tuning, and performance validation to the xDeepFM recommender, which is composed of an embedding layer, a compressed interaction network (CIN), a deep neural network (DNN), and a linear network. 

The input features are in FFM format, including the buyer's address, total amount, collection slug, etc. After the training stage, the trained model is stored on cloud where the server loads from. 

\begin{figure}[h]
  \centering
  \includegraphics[width=5.5cm]{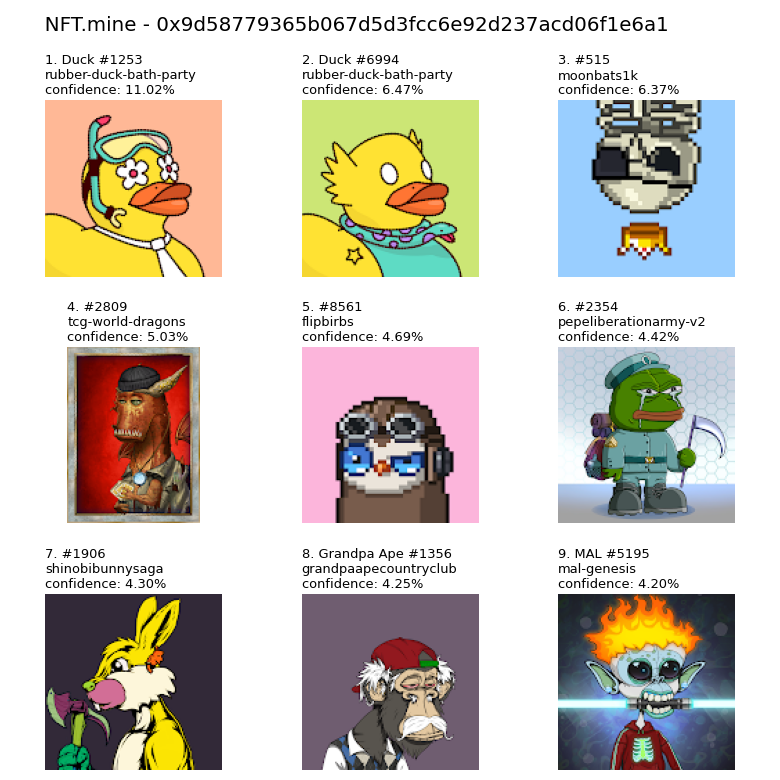}
  \caption{Recommended NFTs in all collections}
  \label{fig:all-nft}
\end{figure}

\begin{figure}[h]
  \centering
  \includegraphics[width=5.5cm]{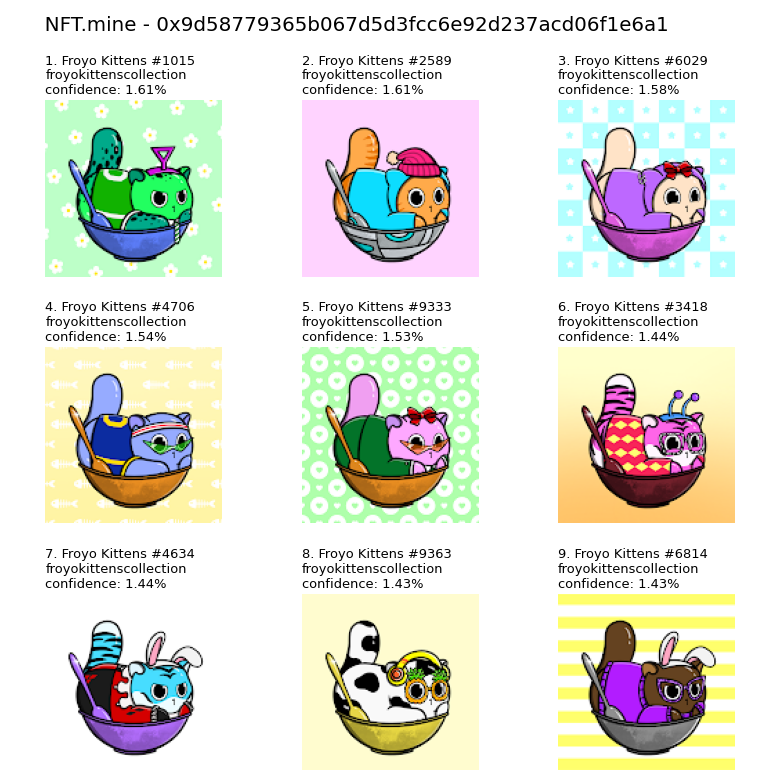}
  \caption{Recommended NFTs in one collection}
  \label{fig:col-nft}
\end{figure}

The server module receives user information as its input and outputs the recommended NFTs. The recommended NFTs are inferred based on the trained model. NFT.mine pairs users to NFTs and inputs these pairs into xDeepFM to obtain a series of probabilities. The NFTs are sorted by probabilities and NFT.mine returns the top-K recommended NFTs (Fig.\ref{fig:all-nft}).

If a user only wants to get the recommended NFTs from a certain collection, he/she could specify the collection and the results only include NFTs from that collection (Fig.\ref{fig:col-nft}).

\section{Evaluation}

In this section, we introduce how to build NFT.mine as an end-to-end recommender system, including environment setup, dataset preparation, training and inference setup, and recommendation performance.

\subsection{Experiment Setup}

We used utilities provided by Microsoft, which support several common tasks for a recommender system, including loading datasets in different formats for different algorithms, splitting training and test datasets, and evaluating model outputs. We used TensorFlow 2.8.0 version for experiments.

\subsubsection{Asset-based and Collection-based Datasets}

For dataset preparation, we used two different methods to split raw data into an asset-based dataset and a collection-based dataset. For asset-based dataset, we grouped data by \texttt{asset\_name}, so each line represents one kind of NFT asset. For collection-based dataset, we grouped data by \texttt{collection\_slug}, so each line represents one kind of NFT collection.

Because xDeepFM receives input data in FFM format: <label> <field\_id>:<feature\_id>:<feature\_value>, before feeding data into NFT.mine, we first converted raw data into FFM format by an open-source libffm tool.

\subsubsection{Training and Inference}

After dataset preparation and format conversion, we split both asset-based and collection-based datasets into training (90\%), validation (10\%), and test (10\%) sets. Then we ran NFT.mine on these datasets. In addition, We implemented baseline models, including linear regression, decision tree, and random forest using sklearn and ran baseline models on the raw dataset.

\subsection{Experiment Results}

\begin{table}
  \caption{Performance of baseline models vs NFT.mine}
  \label{tab:commands}
  \begin{tabular}{cll}
    \toprule
    Model & AUC & Logloss\\
    \midrule
    \texttt{NB} & 0.9360 & 0.1908 \\
    \texttt{RF} & 0.9429 & 0.4747 \\
    \texttt{LR} & 0.9708 & 0.3275 \\
    \texttt{xDeepFM} & 0.9943 & 0.0532 \\    
    \bottomrule
  \end{tabular}
  \label{table:exp}
\end{table}

Table \ref{table:exp} shows AUC and Logloss for baseline models versus xDeepFM. Among all baseline models, the best AUC is 97\% and Logloss is 0.19. For xDeepFM, AUC reaches 99\% and Logloss drops to 0.053. During the experiments, the training loss dropped to almost 0 after epoch 10, evaluation loss dropped to almost 0 after epoch 6, and AUC climbed up to around 1 after epoch 4.

Based on the experiment results, we succeeded in improving our recommender's accuracy by 5\% more compared with traditional approaches.

\section{Conclusion}

NFT.mine is a pioneering work in NFT recommendation. We built an end-to-end NFT recommender system, which pulls the latest data from OpenSea, performs exploratory data analysis, executes model training and inference, and outputs personalized NFT recommendations. NFT.mine achieves an AUC of 99.4\% and Logloss of 0.05. 

\bibliographystyle{ACM-Reference-Format}
\bibliography{sample-base}

\end{document}